\documentclass[aps,twocolumn,preprintnumbers]{revtex4-2}

\usepackage{graphicx}  
\usepackage{subfigure}
\usepackage{multirow}
\usepackage{comment}
\usepackage{fancyhdr}
\usepackage{longtable}
\usepackage{parskip}
\usepackage[T1]{fontenc}
\usepackage{dcolumn}   
\usepackage{xcolor}
\usepackage{bm}        
\usepackage{amsfonts}  
\usepackage{amsmath}   
\usepackage{amssymb}   
\usepackage{appendix}

\begin{document}

\title{Static horizons in cosmology}
\date{\today}
\author{Ida M. Rasulian\footnote{Corresponding author.}, Amjad Ashoorioon}

\affiliation {\it School of Physics, Institute for Research in Fundamental
Sciences (IPM), \\ P.O.Box 19395-5531, Tehran, Iran}
\email{idarasulian@ipm.ir (Corresponding author.), amjad@ipm.ir}

\begin{abstract}Although previous results have ruled out the possibility of a static event horizon in cosmology, we present black hole and white hole metrics that retain static event horizons while reproducing cosmological behavior at large distances. Using an appropriate coordinate choice, we demonstrate that a static event horizon can exist in a cosmological setting without introducing curvature invariant singularities at the horizon. The resulting metric reduces to the Schwarzschild–de Sitter solution when the Hubble parameter is constant. We find that white hole metrics in an expanding universe (or black holes in a contracting universe) are significantly easier to construct, as a black hole in an expanding cosmology requires the velocity function to change sign. Consequently, this work initially examines white holes in expanding cosmologies as a foundation for subsequent analysis of black holes in expanding universes. In later sections, we investigate scenarios involving a white hole coupled with cosmological matter, as well as a white hole with both matter and a cosmological constant. Assuming the pressure component takes its cosmological value, we show that the physical radius of the apparent horizon can asymptotically approach a constant value at late times. This metric avoids pathologies such as a singular horizon in the limit of a vanishing Hubble parameter. Finally, we analyze the realistic case of a black hole embedded in pressureless cosmological matter with and without a cosmological constant and explore its properties. We specifically show that the velocity function can become zero and change sign in the vicinity of a black hole. This means we can smoothly transition from an expanding cosmological phase with a positive velocity function to a contracting black hole phase with a negative velocity function.
\end{abstract}


\maketitle 

\section{Introduction}
Black holes are an indispensable part of any textbook on general relativity, which has established the fundamental mathematical tools in our understanding of the universe as a whole, or in other words, cosmology. Yet, a consistent black hole solution (with or without a static horizon) that can be embedded in cosmology is still a matter of debate (see, for instance, \cite{McVittie, Einstein:1945id, Kaloper, Davidson, Faraoni, Gaur, Dahal, Ion, Poplawski, Farrah, Cadoni1, Cadoni2, Davidson}). Indeed, some arguments have been presented that forbid the existence of black holes with static event horizons, independent of the underlying gravitational theory \cite{Faraoni}.
Our main initiative in this work is twofold. The possibility of static event horizons in cosmology and resolving a common misunderstanding about black holes in a matter-dominated universe, namely the Lemaitre-Tolman-Bondi type metrics.

In the first part of this work, we propose a way out of the no-go results and show that it is possible to find metrics that behave like a black hole close to the horizon and tend to standard cosmology away from it, while keeping the horizon static. To this end, in Section II, we use the Painlevé-Gullstrand coordinates for both the black hole and cosmological metrics. Due to the simplicity of deriving a white hole in expanding cosmology (or similarly, a black hole in contracting cosmology), we first consider this case and study the possibility of a static horizon. In order to achieve a static horizon, we drop one of the assumptions that led to McVittie's metric. We let $p_\perp$ be different from $p_r$ for radial distance $r$ small compared to the cosmological horizon (i.e., in the vicinity of the white hole). Assuming our gravitational theory is general relativity, the energy density everywhere is the same as the cosmological value, and the other components also tend to cosmological values for $\frac{l_1}{r}\ll 1$, where $l_1$ is the static white hole horizon radius. This metric satisfies the null convergence condition, i.e., the null energy condition (NEC) outside the horizon, but it violates it inside the white hole horizon. To resolve this issue, we drop another one of our assumptions, namely, we allow the energy density to be different from the FLRW one, and show that it is, in principle, possible to have a metric that satisfies NEC inside the white hole horizon as well. We then consider the case of a black hole in expanding cosmology that requires a sign change in the velocity function $h$ by considering a more generic metric ansatz.

In the second part of this work (Section III), we turn to the second initiative of this work, which concerns black holes in a matter-dominated universe. Having learned from the first section, we note that embedding a black hole in an expanding cosmology requires a sign change in the velocity function in PG coordinates. In LTB coordinates, this means we need to have a sign change in the derivative of the angular distance function, namely, in our notation $\dot{f}$ should become zero and change sign. To the best of our knowledge, this point is missing from the literature on LTB black holes. In this section, we again begin with the unrealistic case of a white hole in expanding cosmology, since we do not need this sign change for embedding a white hole in expanding cosmology. We then move to the black hole case for which a nontrivial curvature term (even in an asymptotically flat universe) is required to make the aforementioned sign change possible. We stress that if we do not have this sign change in the velocity function, we end up with a white hole metric, rather than a black hole.
In connection with the previous part, we argue that we can have asymptotically static horizons in these cases. We conclude in Section IV.

\section{The feasibility of a static horizon}

In this section, we show that it is in principle possible to have a static horizon in cosmology.

\subsection{A naive extension of the Schwarzschild-de Sitter metric}

The Schwarzschild metric can be written in Painlevé-Gullstrand coordinates as \cite{Painleve, Gullstrand} 
\begin{equation}\label{schw}
ds^2=-(1-\frac{l_s}{r})dt^2+2\sqrt{\frac{l_s}{r}}dr dt+dr^2+r^2d\Omega_2^2.
\end{equation}
On the other hand, the cosmological FLRW metric can be written in Painlevé-Gullstrand coordinates as \cite{Gaur2}
\begin{equation}\label{cosm}
ds^2=-(1-\frac{\dot{a}^2}{a^2}r^2)dt^2-2\frac{\dot{a}}{a}r dt dr+dr^2+r^2 d\Omega_2^2.
\end{equation}
where $a(t)$ is the scale factor. 
This coordinate system is especially suitable for writing a metric that interpolates between the cosmological metric and the Schwarzschild metric due to the time-independent radial factor of the angular coordinates.

Abiding by the requirement that any cosmological black hole metric should tend to the Schwarzschild-de Sitter solution, in the limit of constant Hubble parameter, we first consider the Schwarzschild-de Sitter metric in Painlevé-Gullstrand coordinates \cite{Gaur2}
\begin{equation}
    ds^2=-(1-\frac{l_s}{r}-\frac{r^2}{l^2})dt^2-2\sqrt{\frac{l_s}{r}+\frac{r^2}{l^2}}dt dr+dr^2+r^2 d\Omega_2^2,
\end{equation}
and factorize the $g_{tt}$ part to its roots as
\begin{equation}
    1-\frac{l_s}{r}-\frac{r^2}{l^2}=\frac{l_s}{l_1}(1-\frac{l_1}{r})(1-\frac{r}{l_2})(1+\frac{r}{l_1+l_2}).
\end{equation}
Here, $l_1$ and $l_2$ are respectively the black hole and cosmological horizons, which obey
\begin{equation}
    l_s=\frac{l_1 l_2(l_1+l_2)}{(l_1+l_2)^2-l_1 l_2},\qquad l^2=(l_1+l_2)^2-l_1 l_2.
\end{equation}
Then the metric is brought to the form
\begin{multline}
    ds^2=-\frac{l_s}{l_1}\Big(1-\frac{l_1}{r}-\frac{r^2-l_1^2}{l_2(l_1+l_2)}\Big)dt^2\\-2\sqrt{\frac{l_s}{l_1}\frac{l_1}{r}+\frac{r^2}{l^2}}dt dr+dr^2+r^2 d\Omega_2^2.
\end{multline}
Relating $l_2$ and $l_s$ to $l_1$ and $l$ we can write this as
\begin{multline}\label{dssch}
    ds^2=-\big(1-(\frac{l_1}{l})^2\big)\Big(1-\frac{l_1}{r}-\frac{r^2-l_1^2}{l^2-l_1^2}\Big)dt^2\\-2\sqrt{\big(1-(\frac{l_1}{l})^2\big)\frac{l_1}{r}+\frac{r^2}{l^2}}dt dr+dr^2+r^2 d\Omega_2^2.
\end{multline}
which is useful for our purposes since the dependence on the two horizons is explicit, and they are written in a disentangled fashion.

We should note that due to a negative $g_{tr}$ component, this metric in fact represents a white hole in these coordinates \cite{Volovik}.
Using this analogy, we propose a metric for a spherically symmetric white hole with a static horizon in cosmology as 
\begin{multline}\label{metric}
    ds^2=-(1-\frac{l_1^2 \dot{a}^2}{a^2})\Big(1-\frac{l_1}{r}-\frac{r^2-l_1^2}{\frac{a^2}{\dot{a}^2}-l_1^2}\Big)dt^2\\-2\sqrt{\big(1-\frac{l_1^2 \dot{a}^2}{a^2}\big)\frac{l_1}{r}+\frac{r^2\dot{a}^2}{a^2}}dt dr+dr^2+r^2 d\Omega_2^2.
\end{multline}

The metric \eqref{metric} can be written in a more suggestive form as
\begin{equation}\label{mets}
    ds^2=-(1-h^2)dt^2-2h dt dr+dr^2+r^2 d\Omega_2^2.
\end{equation}
where 
\begin{equation}
    h=\sqrt{\frac{l_1}{r}+\frac{r^3-l_1^3}{r}\frac{\dot{a}^2}{a^2}}.
\end{equation}

This metric has the property that the white hole and cosmological horizons are detached, and it has a static horizon at $r=l_1$. We show in section \ref{secpr} that this is both an apparent and a static horizon.
Also, if we replace $\frac{a}{\dot{a}}$ with the de Sitter length $l$ for the cosmological constant, we recover the Schwarzschild-de Sitter metric.

In the large $r$ limit, the metric tends to the cosmological metric \eqref{cosm} with $t$ the FLRW time. In the small $r$ limit, the metric tends to the Schwarzschild metric \eqref{schw} with Painlevé-Gullstrand time. We note that using this metric, the FLRW time is compared with Painlevé-Gullstrand time in the Schwarzschild metric and not the static Schwarzschild time.

Since the horizon is a little obscure in these coordinates, we can change variables to remove the $g_{tr}$ part of the metric. Writing $t=\gamma(\tau,r)$ the $g_{tr}$ component of the metric becomes
\begin{equation}
    g_{tr}=-2\dot{\gamma}((1-h^2)\gamma'+h).
\end{equation}
Therefore, to remove this, we choose $\gamma$ such that 
\begin{equation}
    \gamma'=-\frac{h}{1-h^2}.
\end{equation}
Then the metric becomes 
\begin{equation}
    ds^2=-(1-h^2)\dot{\gamma}^2 d\tau^2+\frac{dr^2}{1-h^2}+r^2 d\Omega_2^2.
\end{equation}
This metric has a static horizon at $r=l_1$ where $h=1$ independent of $\tau$.

We can see that the Ricci scalar and curvature invariants like $\mathcal{R}^{\mu\nu}\mathcal{R}_{\mu\nu}$ and $\mathcal{R}^{\mu\nu\rho\sigma}\mathcal{R}_{\mu\nu\rho\sigma}$ remain finite at the horizon.
In particular, we have at the horizon $r=l_1$
\begin{equation}
    \mathcal{R}|_{r=l_1}=-6\,l_1\frac{\dot{a}^3}{a^3}+6\,l_1 \frac{\dot{a}\ddot{a}}{a^2}+12\frac{\dot{a}^2}{a^2}.
\end{equation}
For visualization, we sketch the Ricci scalar, $\mathcal{R}^{\mu\nu}\mathcal{R}_{\mu\nu}$, $\mathcal{R}^{\mu\nu\rho\sigma}\mathcal{R}_{\mu\nu\rho\sigma}$ as a function of $r$ for fixed $l_1=1$ and $t=10$ for both radiation and matter-dominated cases in Figure 1.
\begin{figure}
    \centering
    \includegraphics[width=0.7\linewidth]{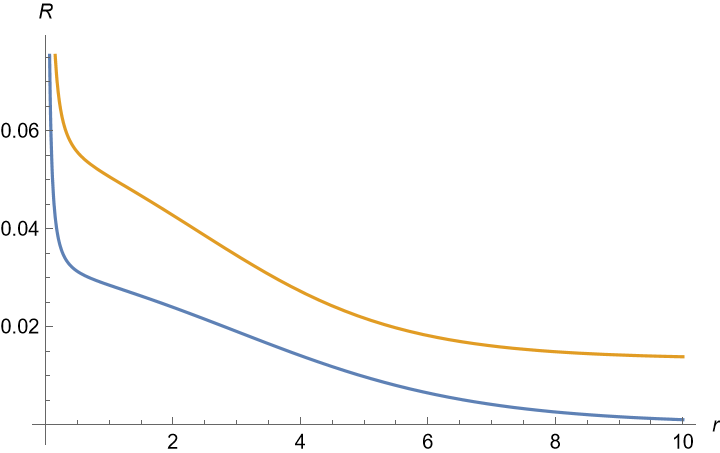}
    \includegraphics[width=0.7\linewidth]{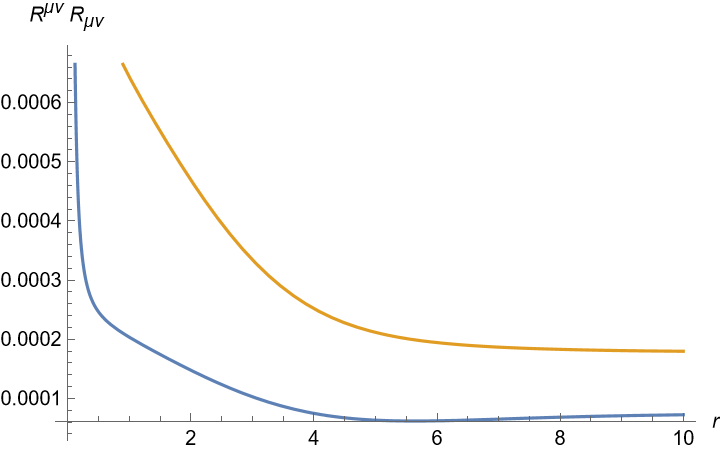}
    \includegraphics[width=0.7\linewidth]{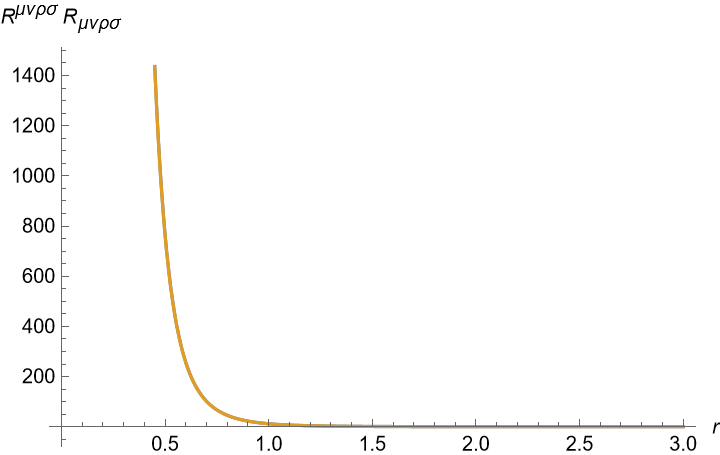}
    \caption{The Ricci scalar, $R^{\mu\nu}R_{\mu\nu}$, $R^{\mu\nu\rho\sigma}R_{\mu\nu\rho\sigma}$ as a function of $r$, for $l_1=1$ and $t=10$, with blue and orange curves related to radiation and matter respectively. In the last plot, radiation and matter curves are almost coincident. These plots represent the observation that this metric has a singularity only at $r=0$ and not at the event horizon.}
    \label{fig:Nonsingular}
\end{figure}

Another interesting property is that $T^t_r$, which we find from requiring Einstein's equation, is zero. In other words, there is no flux of energy in the radial direction. This is the requirement also present in McVittie's solution.
On the other hand, the large $r$ limit of the stress tensor tends to the cosmological values.

Explicitly we have
\begin{equation}
    G^t_t=-3\frac{\dot{a}^2}{a^2},\quad \quad G^t_r=0,
\end{equation}
\begin{equation}
    G^r_t=-2\,r\big(1-\frac{l_1^3}{r^3}\big)\Big(\frac{\dot{a}^3}{a^3}-\frac{\dot{a}\ddot{a}}{a^2}\Big)=g^{t r}(G^r_r-G^t_t),
\end{equation}
and 
\begin{multline}
G^\theta_\theta=G^\phi_\phi=-3\frac{\dot{a}^2}{a^2}-\frac{l_1(l_1^3-7r^3)}{2r^3 g_{tr}^3}\Big(\frac{\ddot{a}\dot{a}}{a^2}-\frac{\dot{a}^3}{a^3}\Big)\\-\frac{(l_1^3-r^3)(l_1^3-4r^3)}{2r^3 g_{tr}^3}\Big(\frac{\dot{a}^5}{a^5}-\frac{\ddot{a}\dot{a}^3}{a^4}\Big)
\end{multline}
We can see that (assuming the gravitational theory is general relativity) the radial pressure for small $r$ is not equal to the angular pressure, i.e., we do not have $p_{\perp}=p_r$. This shows that the metric close to the black hole is not isotropic at every point. But since the $G^\mu_\nu$ tends to the cosmological value for large $\frac{r}{l_1}$, this property ($p_\perp=p_r$) is restored in this limit.

\subsection{A systematic derivation for the white hole case}

Having found the metric in Eq. \eqref{metric}, we show that it could have been derived more systematically and uniquely. We begin with the Painlevé-Gullstrand-like metric ansatz
\begin{equation}\label{ansatz}
    ds^2=-\alpha(t,r)dt^2-2h(t,r)dt dr+dr^2+r^2 d\Omega_2^2,
\end{equation}
and consider the following assumptions:

$a)$ There is no flux of energy in the radial direction, i.e., we assume $T^t_r=0$.

$b)$ There is a static horizon at $r=l_1$, such that $g^{rr}|_{r=l_1}=0$.

$c)$ We assume the energy density is the same as the FLRW one, i.e., we assume $G^t_t=-3\frac{\dot{a}^2}{a^2}$.

$d)$ The metric at large $r$ tends to the cosmological FLRW metric.

We note that, in contrast to McVittie’s assumptions, we have dropped the condition $p_\perp=p_r$ and instead introduced assumption b.

Using the metric \eqref{ansatz}, we have 
\begin{equation}\label{gtr}
    G^t_r=-\frac{h(\alpha'+2h h')}{r(\alpha+h^2)^2},
\end{equation}
where $'$ denotes derivative with respect to $r$. Requiring this to vanish, considering the first assumption, we find
\begin{equation}\label{gtt1}
    \alpha(t,r)=\gamma(t)-\omega(t,r),
\end{equation}
where we have defined $\omega(t,r)=h^2(t,r)$.
We then consider the second assumption. From this and using Eq. \eqref{gtt1} we require
\begin{equation}
    \omega(t,l_1)=\gamma(t).
\end{equation}

Considering assumption c, we have
\begin{equation}
    G^t_t=-\frac{\omega(t,r)+r\omega'(t,r)}{r^2 \omega(t,l_1)}=-3\frac{\dot{a}^2}{a^2}.
\end{equation}
Integrating we find
\begin{equation}\label{omega}
    \omega(t,r)=\omega(t,l_1)r^2\frac{\dot{a}^2}{a^2}+\frac{c(t)}{r}.
\end{equation}
Considering assumption d we have $\omega(t,l_1)=1$ and from this considering Eq. \eqref{omega} at $r=l_1$, we have 
\begin{equation}
    c(t)=l_1\big(1-\frac{\dot{a}^2}{a^2}l_1^2\big).
\end{equation}
Therefore, with these assumptions, the proposed metric in Eq. \eqref{metric} is unique.

\subsection{Energy conditions}
In this section, we study the energy conditions for the stress tensor corresponding to this metric, assuming the gravitational theory is general relativity.

We begin with the weak energy condition that states for every timeline vector $k^\mu$, we should have 
\begin{equation}
    T_{\mu\nu}k^\mu k^\nu\ge 0
\end{equation}
Considering the case where only $k^t=1$ and $k^r$ are nonzero, we have for the vector $k^\mu$ the inequality
\begin{equation}
    g_{tt}+2g_{tr}k^r+g_{rr}(k^{r})^2<0
\end{equation}
which using $g_{tt}=-1+g_{tr}^2$ leads to
\begin{equation}
    -1-g_{tr}<k^r<1-g_{tr}.
\end{equation}
Using these marginal values of $k^r$, we find that in both cases
\begin{equation}
    T_{\mu\nu}k^\mu k^\nu=\frac{2(l_1^3-r^3)(\dot{a}^2-a\ddot{a})}{r^2 a^3 g_{t r}}.
\end{equation}
For a typical power-law scale factor, this is always positive outside the horizon and negative inside the horizon (since $g_{tr}<0$). Therefore, assuming the gravitational theory is general relativity, the stress tensor corresponding to this metric does not satisfy the weak and null energy conditions inside the event horizon.

Considering the strong energy condition, we find that it is violated in the vicinity of the white hole (see figure 2).

\begin{figure}
\includegraphics[width=0.7\linewidth]{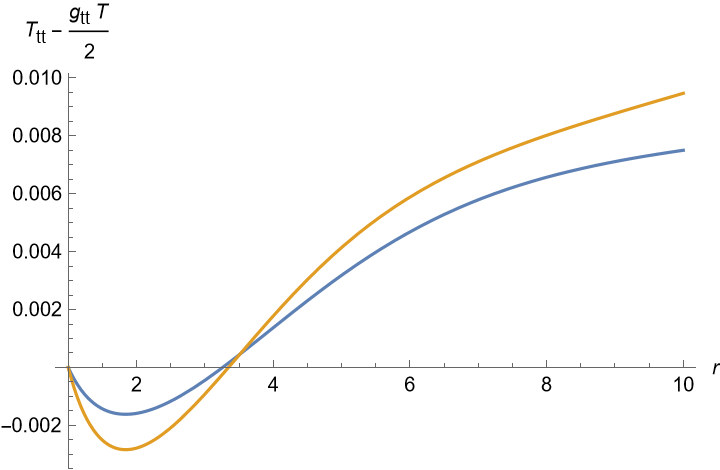}
    \caption{Here we have plotted $T_{tt}-\frac{1}{2}g_{tt} T$ as a function of $r$. We have chosen $l_1=1$ and $t=10$. The blue and orange curves, respectively, correspond to radiation and matter-dominated scale factors. The strong energy condition is violated in the white hole's vicinity.}
    \label{fig:strong-energy}
\end{figure}

\subsection{NEC and static horizon}
Here, we drop assumption c mentioned in the previous section to see if we can find a metric with a static horizon that respects NEC. 
We begin with the metric 
\begin{equation}
    ds^2=-\alpha(t,r)dt^2-2h(t,r)dt dr+dr^2+r^2 d\Omega_2^2.
\end{equation}
We still have \eqref{gtt1}, which considering assumption d takes the form
\begin{equation}
    \alpha(t,r)=1-h^2(t,r).
\end{equation}
We then consider NEC. It can be shown that NEC requires $\dot{h}(t,r)<0$ everywhere, (as we have $T_{\mu\nu}k^\mu k^\nu=-2\frac{\dot{h}}{r}$). Since we assume to have a static horizon at some $r=l_1$ (for which $h(t,l_1)=1$), at the horizon $\dot{h}$ becomes zero and the only way it does not become positive below the horizon is to have $\dot{h}'(t,l_1)=0$ at the horizon. 

One possible metric which respects NEC and assumptions a, b, and d has
\begin{equation}\label{nulec}
    h(t,r)=\sqrt{\frac{l_1}{r}+r^2(1-\frac{l_1^3}{r^3})\frac{\dot{a}^2}{a^2}+l_1^2(1-\frac{l_1^3}{r^3})(\frac{1}{l_0^2}-\frac{\dot{a}^2}{a^2})}.
\end{equation}
Here $l_0$ is related to the cosmological constant $\Lambda=\frac{3}{l_0^2}$.

We have 
\begin{equation}
\dot{h}=-\frac{(l_1-r)^2(l_1+r)(l_1^2+l_1 r+r^2)}{r^3 h}\big(\frac{\dot{a}^3}{a^3}-\frac{\dot{a}\ddot{a}}{a^2}\big),
\end{equation}
which is negative for all $r$.

\subsection{The feasibility of a static horizon, the black hole case}
In this section, we consider a more generic form of the metric that can represent a black hole in cosmology with a static horizon. For this, we consider the metric ansatz \cite{Volovik}
\begin{equation}
    ds^2=-\big(N^2-g_{rr}h^2\big)dt^2+g_{rr}dr^2+2g_{rr}h\,dt dr+r^2 d\Omega_2^2,
\end{equation}
We assume $N^2=\frac{1}{g_{rr}}=\lambda(t,r)$.

Considering the assumption that $G^t_t=-3H^2$, which is its cosmological value, we can write
\begin{equation}
    \partial_r(\frac{r h^2}{\lambda}+r-r\lambda)=H^2 \partial_r(r^3),
\end{equation}
\begin{equation}
    h^2=\lambda(\lambda-1+H^2 r^2+\frac{\alpha(t)}{r})
\end{equation}
Using the requirement that we have a static horizon at $r=l_1$, we find
\begin{equation}
    g^{rr}=\lambda-\frac{h^2}{\lambda}|_{r=l_1}=0.
\end{equation}
From this we have $\alpha(t)=l_1(1-H^2 l_1^2)$ and
\begin{equation}\label{bhm}
    h^2=\lambda\big(\lambda-1+H^2 r^2+\frac{l_1}{r}(1-H^2 l_1^2)\big).
\end{equation}
In accordance with \cite{Volovik}, we can choose 
\begin{equation}
    \lambda=1-3(\frac{l_s H}{2})^{2/3}, \qquad l_s=l_1(1-H^2 l_1^2).
\end{equation}
Then we have 
\begin{equation}
    h^2=(1-\lambda)\lambda(r-r_0)^2\frac{r+2r_0}{3\,r \,r_0^2}, \qquad r_0=(\frac{l_s}{2H^2})^{1/3}.
\end{equation}
Therefore, we can choose different signs of the square root for $r>r_0$ and $r<r_0$.

 Again, one can show that this result violates NEC. Similar to the white hole case, we can drop the assumption that the energy density is equal to the cosmological value and find a metric that also satisfies the NEC.

\subsection{Horizon is apparent and event}\label{secpr}
In this section we show that the white/black hole horizons for the metrics described here are apparent and event horizons.
We begin with \eqref{mets}. One can see that the surface $h=1$ in this space-time describes an apparent horizon. To see this we consider outgoing null rays described with the vector
\begin{equation}
    k_\mu=\{h-1,-1,0,0\},
\end{equation}
This vector deviates from being a geodesic by a factor. We have
\begin{equation}
    k^\alpha_{;\beta}k^\beta=h'\,k^\alpha.
\end{equation}
We introduce a compensating factor $e^{-\Gamma}$ such that $k^\alpha \partial_\alpha \Gamma =h'$. Then $l^\mu=e^{-\Gamma}k^\mu$ \cite{toolkit} satisfies the geodesic equation and we find the expansion as
\begin{equation}
    \theta=l^\alpha_{;\alpha}\propto k^\alpha_{;\alpha}-h'=\frac{2}{r}(h-1).
\end{equation}
Therefore, we see that the expansion changes sign at $h=1$, which takes place at a constant $r=l_1$.

To see whether this is also an event horizon, we first note that this surface is null, stationary, and is an apparent horizon. On the other hand, if we consider the Raychaudhuri's equation
\begin{equation}\label{varthet}
    \frac{d\theta}{d\lambda}=-B^{\alpha\beta}B_{\alpha\beta}-R_{\alpha\beta}k^\alpha k^\beta,
\end{equation}
the second can be shown to vanish at $h=1$, ($r=l_1$) from Einstein's equation. This is because we have
\begin{equation}
    T_{\alpha\beta}k^\alpha k^\beta=-\frac{2}{r}\dot{h},
\end{equation}
which vanishes at $r=l_1$. 

The first term in \eqref{varthet} also vanishes at this surface. We have
\begin{equation}
    B_{\alpha\beta}=k_{\alpha ;\beta}, \qquad B^{\alpha\beta}B_{\alpha\beta}=\frac{2}{r^2}(h-1)^2.
\end{equation}
Therefore, the expansion and its variation at $h=1$ both vanish, and this surface satisfies the characteristics of an event horizon. 

Globally, one can show that this surface is indeed the boundary of the causal future of past null infinity (since we are dealing with a white hole).

Similar considerations apply to the metric \eqref{nulec}. For the black hole case defined as \eqref{bhm}, we repeat the above calculation and find
\begin{equation}
    k_\mu=\{-\lambda+h,1,0,0\},
\end{equation}
\begin{equation}
    k^\alpha_{;\beta}k^\beta=(\lambda'-h')k^\alpha.
\end{equation}
Defining $l^\alpha=e^{-\Gamma}k^\alpha$ such that $k^\alpha \partial_\alpha \Gamma=\lambda'-h'$,
the expansion is
\begin{equation}
    \theta=l^\alpha_{;\alpha}-\lambda'+h'=-\frac{2}{r}(h-\lambda),
\end{equation}
which changes sign at $h=\lambda$, which is the position of the horizon. Therefore, this horizon is an apparent horizon. On the other hand, we have
\begin{equation}
    T_{\alpha\beta}k^\alpha k^\beta=\frac{2}{r}(\dot{h}-\dot{\lambda}),
\end{equation}
which is zero at $r=l_1$ from which $h=\lambda$.

We also have
\begin{equation}
    B_{\alpha\beta}=k_{\alpha ;\beta}, \qquad B^{\alpha\beta}B_{\alpha\beta}=\frac{2}{r^2}(h-\lambda)^2.
\end{equation}
Therefore, the expansion and its variation at $h=\lambda$ ($r=l_1$) both vanish, and this surface satisfies the characteristics of an event horizon. 

\section{The loopholes in previous non-existence proofs}

In this section, we review the arguments in \cite{Faraoni} and mention the loopholes in each argument in each case. The first argument is based on the assumption that all such metrics can be written in the form
\begin{equation}\label{f1}
    ds^2=-T^2(t,R)dt^2+a^2(t,R)(dR^2+R^2 d\Omega_2^2).
\end{equation}
If we transform our proposed metric to such coordinates, using a transformation $r=f(t,R)$ such that $h=-\dot{f}$, it will be of the form
\begin{equation}
    ds^2=-N^2 dt^2+\frac{f'^2}{\lambda}dR^2+f^2 d\Omega_2^2.
\end{equation}
The special example of this is the Schwarzschild metric in the Lemaitre coordinates, obtained by assuming $f=\frac{3}{2}l_s^{\frac{1}{3}}(R-t)^{\frac{2}{3}}$, which cannot be written in the form \eqref{f1}. On the other hand, following an argument similar to \cite{Faraoni}, we find that at the proposed $f=\text{const.}$ location of the horizon, we need to have
\begin{equation}\label{temp}
    \dot{f}^2=\lambda N^2.
\end{equation}
One may naively think, since the right-hand side of this equation is always positive, how can it be that this equality holds at some constant $f$? The simplest counterexample is again the Schwarzschild metric in Lemaitre coordinates, for which the horizon is at $f=l_s$ or, in these coordinates, $R-t=l_s$. This satisfies \eqref{temp} with $\lambda=N=1$.

We then move to the second argument of \cite{Faraoni} that describes a contradiction in assuming a static horizon by showing that the radial null and time-like geodesics essentially freeze at the would-be static horizon. We note that this behavior is what an asymptotic (accelerated) observer sees when looking at in-falling matter. Even for a pure black hole, with mass $m$, whose approximate near-horizon metric is
\begin{equation}
    ds^2=-\frac{x^2}{4}dt^2+4m^2(dx^2+d\omega_2^2),
\end{equation}
the radial null geodesics are
\begin{equation}
    u^\mu=\{\pm\frac{4}{x},\frac{1}{m},0,0\},
\end{equation}
whose velocity $\frac{u^1}{u^0}$ tends to zero as they approach the horizon at $x=0$.

The arguments regarding the divergence of the Ricci scalar at the static horizon are explicitly ruled out by our choice of metric, as can be seen in Fig. \ref{fig:Nonsingular}. There is no divergent behavior for curvature invariants here.

Regarding the Hawking temperature, one can expect that $a(t,R)$ varies such that close to the black hole region the black hole temperature remains approximately constant. As explained in the next section, the behavior of $a(t,R)$ is different in the black hole region (contracting) and the cosmological region (expanding).

\section{The case of constant equation of state}
In this section, we consider the possibility that the metric respects the cosmological stress tensor equation of state.
\subsection{The white hole case}
Here we use the metric ansatz \eqref{ansatz}.
Assuming $T^t_r=0$, from \eqref{gtr} we find
\begin{equation}
    \alpha(t,r)=\gamma(t)-h^2(t,r).
\end{equation}
We do not consider any further assumptions and continue our analysis for the case of a constant equation of state.

We assume $p_\perp=p_r=w \rho$. Assuming $G^r_t=g^{tr}(G^r_r-G^t_t)$ \footnote{This follows from $G^t_r=0$ and $G^{tr}=G^{rt}$.}, we have from Einstein equations
\begin{equation}
   G^r_t=\frac{h}{r\gamma^2}(2\gamma \dot{h}-h \dot{\gamma})=-\frac{h}{\gamma^2}\kappa_4(1+w) \rho,
\end{equation}
and therefore 
\begin{equation}\label{equ}
    \frac{1}{r}(2\gamma \dot{h}-h\dot{\gamma})=-\kappa_4 (1+w)\rho.
\end{equation}
On the other hand from $G^t_t=-\kappa_4 \rho$ and Eq. \eqref{equ}, we find
\begin{equation}\label{first}
    h^2+2r h h'+\frac{2r \gamma^2 \dot{h}-r h \gamma \dot{\gamma}}{1+w}=0.
\end{equation}
We also have by considering $G^r_r=\kappa_4 w \rho$,
\begin{equation}\label{second}
    h^2+2r h h'+\frac{2r \dot{h}-r h\frac{\dot{\gamma}}{\gamma}}{1+w}=0.
\end{equation}
Using \eqref{first} and \eqref{second} we find 
\begin{equation}
    (2r \dot{h}-r h \frac{\dot{\gamma}}{\gamma})(\gamma^2-1)=0.
\end{equation}
So we either have $\gamma^2=1$ or $2\frac{\dot{h}}{h}=\frac{\dot{\gamma}}{\gamma}$.

The second possibility requires $h^2=c(r)\gamma(t)$. Using \eqref{first} again we find $c(r)=\frac{l_s}{r}$ where $l_s$ is a constant length and the metric takes the form
\begin{equation}
    ds^2=-\gamma(t)(1-\frac{l_s}{r})dt^2-2\sqrt{\frac{l_s}{r}\gamma(t)}dt dr+dr^2+r^2 d\Omega^2,
\end{equation}
which corresponds to the Schwarzschild white hole metric written in Painlevé–Gullstrand coordinates (as can be seen by introducing a new time coordinate $t'$, defined by $dt'=\sqrt{\gamma(t)}dt$). Therefore, in constructing a cosmological white hole metric, we shall henceforth set $\gamma(t)=1$.

Assuming $\gamma(t)=1$, we have from the $G^0_0$ component of Einstein equations and \eqref{equ}
\begin{equation}\label{gtt}
    h^2+2r h h'+\frac{2}{1+w}r \dot{h}=0.
\end{equation}
Considering the $G^\theta_\theta$ component we find
\begin{equation}\label{gtheta}
    \partial_r(r\dot{h}+r h h'+\frac{h^2}{2})=-\kappa_4 r w \rho=\frac{1}{r}(h^2+2r h h'+2r \dot{h}),
\end{equation}
where we used the $G^r_r$ component as well.
In the following, we consider the case of $\omega=0$ and $\omega\neq 0$ separately.

\subsubsection{The case of $w\neq 0$}
In this case, from \eqref{gtheta} we find
\begin{equation}\label{eq}
    h^2+2r h h'+2r \dot{h}=\frac{2w}{1+w}\dot{\zeta}(t)r^2.
\end{equation}
Here $\zeta(t)$ is some function of $t$, and we have chosen the coefficient and a derivative with respect to $t$ on the right-hand side for later convenience.

Subtracting \eqref{gtt} from \eqref{eq}, we further find
\begin{equation}
    h=\zeta(t)r+\eta(r).
\end{equation}
Using \eqref{gtt} again we find
\begin{multline}
   3\zeta^2(t)r^2+4r \zeta(t)\eta(r)+2r^2 \zeta(t)\eta'(r)\\+\frac{2}{1+w}r^2
\dot{\zeta}+\eta^2(r)+2r \eta(r)\eta'(r)=0.
\end{multline}
The only nontrivial solution of this equation which is of cosmological interest is $\eta(r)=0$, $\zeta(t)=\frac{2}{3(1+w) t}$. Therefore, we do not have a cosmological solution that satisfies $p_r=p_\perp=w \rho$ for $w\neq 0$.

\subsubsection{The case of matter with $w=0$}
To simplify our analysis, we change the coordinates to a semi-homogeneous form by setting $r= f(t,R)$, demanding the transformed metric to be diagonal by setting $h=\dot{f}$. Therefore, the metric becomes simply
\begin{equation}
    ds^2=-dt^2+f'^2 dR^2+f^2 d\Omega_2^2,
\end{equation}
where $f'=\partial_R f$. This metric is a special case of Lemaitre-Tolman-Bondi (LTB) type metrics with vanishing curvature that are well studied in the literature (see, for instance, \cite{Bondi, Tolman, Krasinski, Gorini, Joshi, Gao, Kopteva, Kopteva2, Valsberg}). The important observation here is that with a trivial choice for the curvature term, we end up with the white hole solution, since there will be no sign change in the velocity function $h=\dot{f}$. In the following, we present a self-contained treatment of this metric ansatz with a specific matter profile.

We note that if we had $f=R\, a(t)$, this was the FLRW metric. In this coordinate $T^R_t=T^t_R=0$ and the Einstein equations become
\begin{equation}\label{density}
    \frac{\dot{f}^2}{f^2}+\frac{2\dot{f}\dot{f}'}{f f'}=\frac{\partial_R (f \dot{f}^2)}{f^2 f'}=\kappa \,\rho.
\end{equation}
\begin{equation}\label{rpressure}
    \frac{\dot{f}^2}{f^2}+2\frac{\ddot{f}}{f}=-\kappa\, p_R,
\end{equation}
\begin{equation}\label{perpressure}
    \frac{\dot{f}\dot{f}'+f' \ddot{f}+f \ddot{f}'}{f f'}=-\kappa \,p_\perp.
\end{equation}

In this section, we consider the case of a Schwarzschild white hole immersed in cosmological matter. In this case, we have $p_r=p_\perp=0$. From stress-energy conservation, we can find the structure of $\rho$. We have 
\begin{equation}\label{rhocons}
    \frac{\dot{\rho}}{\rho}+2\frac{\dot{f}}{f}+\frac{\dot{f}'}{f'}=0
\end{equation}
and we find
\begin{equation}\label{rho}
    \kappa \rho=\frac{\beta(R)}{f^2 f'}.
\end{equation}
From \eqref{density} we can write
\begin{equation}\label{int}
    f\dot{f}^2=\zeta(R), \qquad \zeta'(R)=\beta(R).
\end{equation}

From \eqref{rpressure} we find
\begin{equation}
    f(t,R)=\zeta(R)^{\frac{1}{3}}\,\big(\frac{3}{2}t-b(R)\big)^{\frac{2}{3}},
\end{equation}
which also satisfies \eqref{perpressure}.

We first re-derive the Schwarzschild metric in these coordinates.
If we only had a point mass contribution to $\rho$, this equation would lead to 
\begin{equation}
\zeta(R)=l_s
\end{equation}
or 
\begin{equation}
f(t,R)=l_s^{\frac{1}{3}}\left(\frac{3}{2}t-b(R)\right)^{\frac{2}{3}}.
\end{equation}
We note that if we choose $b(R)=\frac{3}{2}R$, this will be a white hole metric in Lemaitre coordinates \cite{Lemaitre}.

In our original central coordinate, this means that we had \begin{equation}
h=\dot{f}=\frac{l_s^{1/3}}{\left(\frac{3}{2}t-b(R)\right)^{1/3}}=\sqrt{\frac{l_s}{r}}
\end{equation}
where in the last equality we used $r=f$. We see that this simple case corresponds to the Schwarzschild metric in Painlevé-Gullstrand coordinates. We also note that in these coordinates the singularity is at the surface $\frac{3}{2}t-b(R)=0$. There is freedom in our choice of $b(R)$. 

Now, returning to the case of matter-dominated cosmology, we can add a term related to the cosmological matter density to $\beta(R)$. We note that in the asymptotic case we should have $\kappa_4\beta(R)=\beta_0 R^2$ where $\beta_0$ is a constant. Considering that this is also the case for large $t$ we set $\kappa_4\beta(R)=\beta_0 R^2$ and find
\begin{equation}
    \zeta(R)=l_s+\frac{\beta_0}{3}R^3,
\end{equation}
and 
\begin{equation}
    h=\dot{f}= \frac{\left(l_s+\frac{\beta_0}{3}R^3\right)^{\frac{1}{3}}}{\left(\frac{3}{2}t-b(R)\right)^{\frac{1}{3}}}=\frac{r}{\frac{3}{2}t-b(R)}.
\end{equation}
The apparent horizon is at $h=1$. Therefore we have
\begin{equation}\label{rad}
    r_{H}=\frac{3}{2}t-b(R_H).
\end{equation}
On the other hand, using $r=f$ we have 
\begin{equation}
    r_H=(l_s+\frac{\beta_0}{3}R^3)^{\frac{1}{3}}(\frac{3}{2}t-b(R))^{\frac{2}{3}},
\end{equation}
and using \eqref{rad} we can write
\begin{equation}\label{rad2}
    r_H=l_s+\frac{\beta_0}{3} R_H^3.
\end{equation}
As an example, if we choose $b(R)=\frac{\kappa l_s^4}{R^3}-d$, using \eqref{rad} and \eqref{rad2} we find 
\begin{equation}\label{rad3}
    r_H=l_s+\frac{\frac{\beta_0}{3} \kappa l_s^4}{\frac{3}{2}t+d-r_H},
\end{equation}
with solutions
\begin{eqnarray}\label{rht}
    r_H(t)&=&\frac{1}{4}\bigg(3t+2d+2l_s  \pm \nonumber \\
    &&\sqrt{(3t+2d-2l_s)^2-\frac{16}{3}\beta_0 \kappa l_s^4}\bigg).
\end{eqnarray}
For large $t$ one of these tends to $r_H=\frac{3t}{2}$, which is the cosmological horizon, and the other tends to $r_H=l_s$, which is the white hole horizon. Therefore, for large $t$ we effectively have a static white hole horizon.

We had freedom in our choice of $b(R)$. Requiring that the metric tends to the cosmological metric for large $R$ will force us to choose $b(R)$ such that it tends to a constant for large $R$. 
The surface $\frac{3t}{2}=b(R)$, $r=0$ is where both the cosmological singularity and white hole singularity reside. 

We can prove that generically the resulting apparent horizon in this case will be dynamical (apart from the asymptotic behavior). To see this, we write 
\begin{equation}
    h=\dot{f}=\frac{\zeta^{\frac{1}{3}}}{(\frac{3t}{2}-b(R))^{\frac{1}{3}}}=\frac{r}{\frac{3t}{2}-b(R)}.
\end{equation}
Requiring $h=1$ at the horizon and considering $r=f$, we can write
\begin{equation}
    r_H=\zeta(R_H)=\int^{R_H}dR' \beta(R'),
    \end{equation}
    \begin{equation}
    r_H=\frac{3t}{2}-b(R_H).
\end{equation}
Taking a time derivative, we find
\begin{equation}
    \dot{r}_H=\dot{R}_H \beta(R_H)=\frac{3}{2}-\dot{R}_H b'(R_H).
\end{equation}
From this, we can see that it is impossible to have $\dot{r}_H=0$ exactly \footnote{To have $\dot{r}_H=0$ from the first equality we should either have $\dot{R}_H=0$ which contradicts the second equality or we should have $\beta(R_H)=0$ which again leads to $R_H=const.$ which is a contradiction.}. 
\subsubsection{The case of matter plus cosmological constant}
In this case from $-\kappa_4 p_r=\frac{3}{l^2}$ and Eq. \eqref{rpressure} we find
\begin{equation}\label{acc}
   f(t,R)=\zeta^{\frac{1}{3}}\sinh^{\frac{2}{3}}\left(\frac{3t}{2l}-\frac{b(R)}{l}\right). 
\end{equation}
We again have Eqs. \eqref{rhocons} and \eqref{rho} for the matter part of energy density. From Eq. \eqref{density} we can write
\begin{equation}
  \partial_R(f \dot{f}^2)=
\beta(R)+\frac{3}{l^2}f^2 f'.\end{equation}
 Using Eq. \eqref{acc} we find
\begin{equation}
    \frac{1}{l^2}\zeta'= \beta(R).
\end{equation}
We are again left with two unfixed functions of $R$, namely $b(R)$ and $\beta(R)$. One can use the degree of freedom in choosing the $R$ coordinate to fix one of them. 

If we assume $\beta(R)=\beta_0 R^2$ with $\beta_0$ a constant we find
\begin{equation}\label{cr}
    \zeta(R)=l_s l^2+\frac{\beta_0\, l^2}{3}R^3
\end{equation}
By studying $h=R \,\dot{a}$ one can see that for $\beta_0=0$, which is the case of pure cosmological constant, $l_s$ is the Schwarzschild radius, i.e., for $\beta_0=0$ we have 
\begin{equation}
h|_{\beta_0=0}=\dot{f}=\sqrt{\frac{l_s}{r}+\frac{r^2}{l^2}}.
\end{equation}

We study the apparent horizon, which is located at $h=1$. We have 
\begin{equation}
    h=\dot{f}=\frac{r}{l}\coth\left(\frac{3t}{2l}-\frac{b(R)}{l}\right).
\end{equation}
From this at $h=1$ we find
\begin{equation}
    \sinh^2\left(\frac{3t}{2l}-\frac{b(R)}{l}\right)=\frac{\frac{r_H^2}{l^2}}{1-\frac{r_H^2}{l^2}}.
\end{equation}
Using Eqs. \eqref{acc} and \eqref{cr} and assuming $b(R)$ behaves such that the location of the horizon for large $t$ goes to small $R_H$, we can write for large $t$
\begin{equation}
    r_H\approx (l_s l^2)^{1/3}\Big(\frac{\frac{r_H^2}{l^2}}{1-\frac{r_H^2}{l^2}}\Big)^{1/3},
\end{equation}
and 
\begin{equation}
    \frac{l_s}{r_H}+\frac{r_H^2}{l^2}=1.
\end{equation}
The smallest root of this equation is the white hole horizon.
This means that the location of the white hole horizon for large $t$ tends to its Schwarzschild-de Sitter counterpart.

If we require that the observer is at large $R$, then we should choose $b(R)$ such that the cosmological behavior is dominant for large $R$. The location of singularity is at $\frac{3t}{2}=b(R)$. Choosing as an example $b(R)=\frac{\kappa l_s^4}{R^3}-d$, the location of singularity tends to $R=0$ for large $t$ and to a constant $R$ for small $t$.
\subsection{The Black Hole case}
In this section, we study the case of a black hole immersed in cosmological matter and matter plus cosmological constant. We will see that the solutions that represent black holes are LTB metrics with necessarily non-trivial curvature terms. The latter is a result of requiring a sign change in the velocity function in the vicinity of the black hole. 

\subsubsection{The case of matter with positive curvature}
Beginning from the metric ansatz 
\begin{equation}
    ds^2=-N^2 dt^2+g_{rr}(dr-h dt)^2+r^2 d\Omega_2^2,
\end{equation}
we find after a change of variable $r=f(t,R)$ such that $h=\dot{f}$ the metric
\begin{equation}
    ds^2=-N^2 dt^2+g_{rr}(f')^2dR^2+f^2 d\Omega_2^2.
\end{equation}
Assuming we only have a dust contribution to the stress tensor such that only $T^t_{\,t}=-\kappa_4\rho$ is nonzero, we find from stress tensor conservation
\begin{equation}
    N'=0.
\end{equation}
Therefore, $N$ is only a function of $t$, and by redefining $t$, we can set it to $N=1$.

Requiring $T^t_{\,R}=0$ we further find $\dot{g}_{rr}=0$ and we define $g_{rr}=\frac{1}{1-\omega(R)}$. So far, we see that the metric belongs to the family of LTB metrics (see, for instance \cite{Bondi, Tolman, Krasinski, Gorini, Joshi, Gao, Kopteva, Kopteva2, Valsberg}). Here, we provide a self-contained treatment of this solution with a focus on the sign change in the velocity function. 

From $T^R_{\,R}=0$ we find
\begin{equation}
    \partial_t\big(\omega(R)f+f \dot{f}^2\big)=0,
\end{equation}
On the other hand, from the conservation equation, we can fix the form of $\kappa_4\rho$ to be
\begin{equation}
    \kappa_4 \rho=\frac{\beta(R)}{f^2 f'}
\end{equation}
and from $T^t_{\,t}=-\kappa_4 \rho$ we find
\begin{equation}
    \partial_R\big(\omega(R) f+f \dot{f}^2\big)=\beta(R).
\end{equation}
Therefore, we can write this as
\begin{equation}\label{maineq}
    f \dot{f}^2=\zeta(R)-\omega(R)f
\end{equation}
where we have defined $\zeta(R)$ such that $\zeta'(R)=\beta(R)$.

The general solution to this equation can be written as \cite{Kopteva}
\begin{equation}\label{sol}
    \frac{\zeta}{\omega\sqrt{\omega}}\arcsin{\sqrt{\frac{\omega f}{\zeta}}}-\frac{\sqrt{f(\zeta-\omega f)}}{\omega}=\pm t+b_{\pm}(R).
\end{equation}
We note that \eqref{maineq} requires $f<\frac{\zeta}{\omega}$ for $\omega>0$.

We study the behavior of this solution in different limits. For $f\ll \frac{\zeta}{\omega}$ the solution becomes
\begin{equation}\label{bhf}
    f=\zeta^{1/3}(R)\left(\frac{3}{2}(\pm t+b_{\pm}(R))\right)^{2/3}.
\end{equation}
In the limit where $f\approx \frac{\zeta}{\omega}$ the first term in \eqref{sol} is dominant and the solution takes the form
\begin{equation}
    f=\frac{\zeta}{\omega}\sin^2\big(\frac{\omega^{3/2}}{\zeta}(\pm t+b_{\pm}(R))\big),
\end{equation}
such that at $\pm t+b_{\pm}(R)=\frac{\pi\,\zeta(R)}{2\,\omega^{3/2}(R)}$, $f$ reaches its maximum for a given $R$.

\begin{figure}
    \centering
    \includegraphics[width=0.7\linewidth]{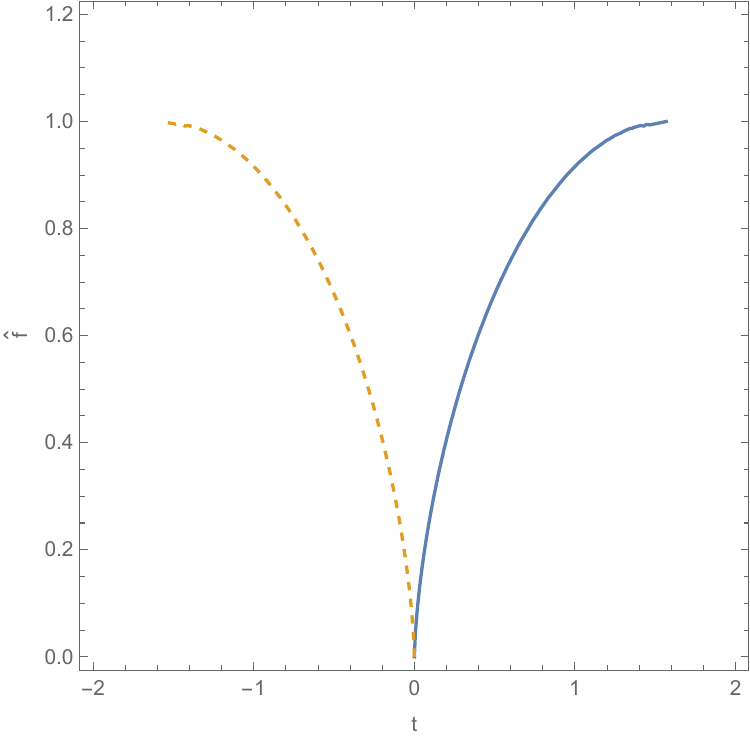}
    \caption{Plot of $\hat{f}$ in terms of $t$ with $b_{\pm}=0$. Here, the plain line represents the $-t$ branch and the dashed line represents the $+t$ branch.}
    \label{fig:bzero}
\end{figure}

\begin{figure}
    \centering
    \includegraphics[width=0.7\linewidth]{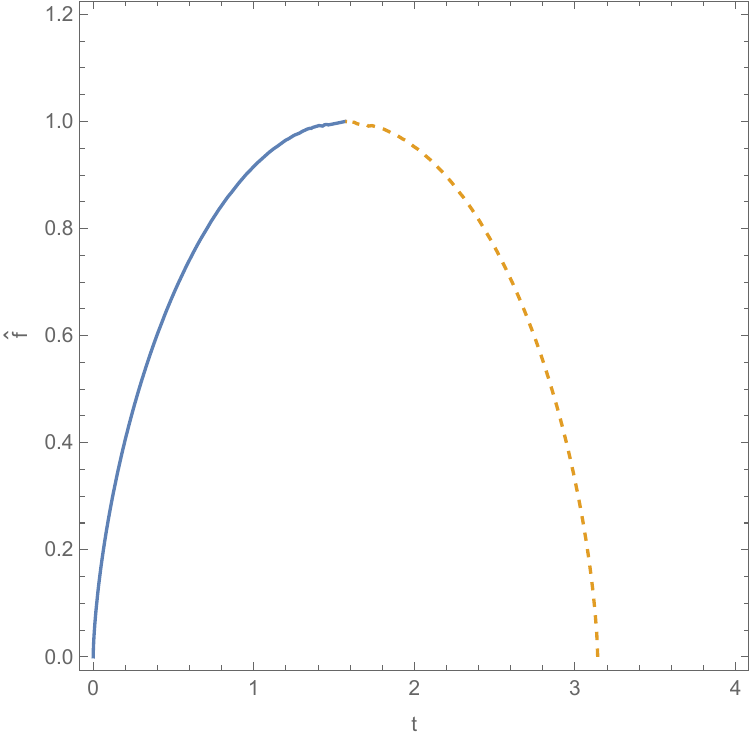}
    \caption{The plot for $\hat{f}$ with $b_{\pm}$ chosen such that we have a smooth transition between the $+$ and  $-$ branches.}
    \label{fig:ba}
\end{figure}

To understand the behavior of the solution we define $f=(\frac{\zeta}{d})^{1/3}\hat{f}$ where $d$ has dimension length and assume $\omega(R)=(\frac{\zeta}{d})^{2/3}$ to find
\begin{equation}
    \hat{f} \dot{\hat{f}}^2=d-\hat{f},
\end{equation}
with solutions
\begin{equation}
   d\arcsin{\sqrt{\frac{\hat{f}}{d}}}-\sqrt{\hat{f}(d-\hat{f})}=\pm t+b_{\pm}(R).
\end{equation}
For a fixed $R$, $b_{\pm}(R)$ are constants. If we set $b_{\pm}=0$, then the solutions are as in Figure \ref{fig:bzero}.
To have a reasonable solution that moves from one solution smoothly to the other, we note that $\hat{f}$ takes its maximum value at $\pm t^c_{\pm}+b_{\pm}=\frac{\pi\, d}{2}$ where $t^c_{\pm}$ are transition times. Therefore, for a smooth transition, we need to have $t^c_-=t^c_+=t^c$ from which we find $b_++b_-=\pi\, d$. For $b_+=0$ and $b_-=\pi\, d$, the plot becomes as in Figure \ref{fig:ba}.
According to this, the behavior of $\hat{f}$ is such that at each value of $R$, for early time ($t<t^c=\frac{\pi\, \zeta}{2\,\omega^{3/2}}$) we have an increasing scale factor reminiscent of an expanding cosmology. After the crossover time, the scale factor becomes decreasing, which is reminiscent of the behavior in the vicinity of a black hole.

Given this intuitive picture, we proceed with a more in-depth analysis of this geometry. Defining $y=\frac{\omega}{\zeta}f$ we can write \eqref{sol} as
\begin{equation}\label{sol1}
    \arcsin \sqrt{y}-\sqrt{y-y^2}=\frac{\omega^{3/2}}{\zeta}(\pm t+b_{\pm}(R)).
\end{equation}
For every value of $R$, there is a crossover time $t_c(R)$ where $\dot{f}$ (and equivalently $\dot{y}$) becomes zero, and we transition from the expanding cosmological phase with $\dot{f}>0$ to a contracting black hole phase with $\dot{f}<0$. At this point, $f$ takes its maximal value $f_c(R)=\frac{\zeta}{\omega}$, or equivalently $y$ becomes $1$ and we have from \eqref{sol1}
\begin{equation}
    \frac{\pi}{2}=\frac{\omega^{3/2}}{\zeta}(\pm t_c(R)+b_{\pm}(R)).
\end{equation}
Summing the two equations we find
\begin{equation}
    b_-(R)+b_+(R)=\frac{\pi\,\zeta}{\omega^{3/2}}.
\end{equation}
In the far $+$ branch, which is pure cosmology to a good approximation, it is consistent to choose $b_+(R)=0$. Then we have
\begin{equation}\label{omega1}
\frac{\omega^{3/2}}{\zeta}=\frac{\pi}{b_-(R)}.
\end{equation}
Then, in the $+$ branch (cosmological expanding realm), we can write
\begin{equation}\label{pluseq}
    \arcsin \sqrt{y}-\sqrt{y-y^2}=\frac{\pi t}{b_-(R)},
\end{equation}
and in the contracting $-$ branch we can write
\begin{equation}\label{minuseq}
     \arcsin \sqrt{y}-\sqrt{y-y^2}=\frac{-\pi t}{b_-(R)}+\pi,
\end{equation}
such that $y$ is only a function of $\frac{t}{b_-(R)}$.

In the deep $+$ region for which $y\ll 1$ we can approximate $y$ and $f$ as
\begin{equation}
    y\approx \left(\frac{3\pi t}{2b_-(R)}\right)^{2/3}, \qquad f\approx \zeta^{1/3}\left(\frac{3t}{2}\right)^{2/3}.
\end{equation}

Using the freedom in defining the $R$ coordinate we choose $\zeta$ as
\begin{equation}
    \zeta(R)=l_s+\frac{\beta_0}{3}R^3.
\end{equation}
We note that the constant shift, $l_s$, is due to the black hole, and the other part is purely cosmological. 

The spatial curvature is proportional to $\frac{\omega}{f^2}$. In order to have negligible curvature for large $f$, we require $\omega(R)$ to be a dimension-less constant that we choose to be $\omega_0=(\beta_0 l_s^2)^{\frac{1}{3}}$ such that for both $l_s=0$ and $\beta_0=0$ we have zero curvature. We note that simply requiring $\omega$ to tend to zero for large (or small) values of $R$ is not enough, since small $R$ for late time can correspond to large $f$. Therefore, to have an asymptotically vanishing spatial curvature, we need to choose a nonzero constant for $\omega_0$ that becomes zero when the black hole or cosmological matter is absent. This constant $\omega$ behaves like a conical singularity at the center of the $R$ coordinate.

According to Eq. \eqref{bhf}, the metric away from the transition point only depends on $\zeta$ and $b_{\pm}$. Therefore, our calculations regarding the apparent horizon remain the same as the white hole case, and we can have an apparent horizon that tends to a constant for large times.

Intuitively, we expect a non-zero flow towards the black hole. Here we show that this is indeed the case in appropriate (black hole) coordinates.

We can write the metric as
\begin{equation}
    ds^2=-dt^2+\frac{(df-\dot{f}dt)^2}{1-\omega}+f^2 d\Omega_2^2.
\end{equation}
Using the equation of motion for $f$, this can be written as
\begin{equation}
    ds^2=-\frac{1-\frac{\zeta}{f}}{1-\omega}dt^2+\frac{df^2}{1-\omega}-\frac{2\dot{f}}{1-\omega}dt df+f^2 d\Omega_2^2.
\end{equation}
For simplicity, we work in the limit where $\omega\approx 0$. Then we can write it as
\begin{equation}
    ds^2=-(1-\frac{\zeta}{f})\big(dt\mp\frac{\sqrt{\frac{\zeta}{f}}}{1-\frac{\zeta}{f}}df\big)^2+\frac{df^2}{1-\frac{\zeta}{f}}+f^2 d\Omega_2^2.
\end{equation}
We note that the infinitesimal time coordinate in the black hole frame is $dT=dt\mp\frac{\sqrt{\frac{\zeta}{f}}}{1-\frac{\zeta}{f}}df$. Therefore, we can write
\begin{equation}
    \rho dt^2=\hat{T}_{TT}dT^2+2\hat{T}_{Tf}dT df+\hat{T}_{ff}df^2,
\end{equation}
where $\hat{T}_{\mu\nu}$ is the stress tensor in the new (black hole) frame. From this, using the equation for $dT$, we can derive
\begin{equation}
    \hat{T}_{TT}=\rho, \qquad \hat{T}_{Tf}=\pm \frac{\sqrt{\frac{\zeta}{f}}}{1-\frac{\zeta}{f}}\rho, \qquad \hat{T}_{ff}=0.
\end{equation}
Therefore, the flow in the black hole frame is nonzero as expected.

\subsubsection{The case of matter with zero curvature}
In this section we consider the case of $\omega=0$ and study the possibility of having a sign change in the velocity function. In this case, the solution to $f$ is known to be 
\begin{equation}
    f=\zeta(R)^{\frac{1}{3}}\big(\frac{3}{2}(\pm t+b(R))\big)^{\frac{2}{3}}.
\end{equation}
The only way to go from a positive branch to a negative branch is if $\zeta(R)\geq 0$ becomes zero at some $R=R_c$.Considering the positivity of $\zeta$, this means $\zeta'$ will become negative for some $R<R_c$. On the other hand, we note that
\begin{equation}
    \kappa \rho=\frac{\zeta'}{f^2 f'}>0,
\end{equation}
and therefore $f'$ should be negative in the same region $R<R_c$. This is related to the concept of shell crossing, which we avoid here \cite{book}.

Since $f'=\zeta'=0$ at $R=R_c$ we need $b'(R_c)=0$. We choose the plus branch for $R>R_c$ and the minus branch for $R<R_c$. We note that to have a finite value for $f$ at $R=R_c$, $b(R)$ should diverge like $\frac{1}{\sqrt{\zeta}}$ at $R=R_c$. On the other hand, in the minus branch the black hole singularity is at $t=b(R)$, which means for $t$ tending to infinity, $R$ will tend to $R_c$. This is a contradiction since we should not have the black hole singularity at the transition point $R=R_c$.

\subsubsection{The case of matter plus cosmological constant}

Similar to the previous section, we can again set $N=1$ and from $T^t_R=0$ we find $\dot{g}_{RR}=0$. From $\kappa_4 T^R_R=-\frac{3}{l_0^2}$ related to the cosmological constant we can write
\begin{equation}
    \partial_t\big(\omega f+f \dot{f}^2-\frac{f^3}{l_0^2}\big)=0.
\end{equation}
On the other hand, from energy-momentum conservation, we find as before
\begin{equation}
    \kappa_4\rho=\frac{\beta(R)}{f^2 f'}.
\end{equation}
From $\kappa_4 T^t_t=-\kappa_4\rho-\frac{3}{l_0^2}$ we find
\begin{equation}
    \omega f+f \dot{f}^2-\frac{f^3}{l_0^2}=\zeta(R),
\end{equation}
where we have $\zeta'(R)=\beta(R)$.

We study this equation in two limits. Close to the black hole, we expect from previous results that $f\ll \frac{\zeta}{\omega}$. In this limit, we have 
\begin{equation}
    f\dot{f}^2-\frac{f^3}{l_0^2}\approx \zeta(R),
\end{equation}
and we find
\begin{equation}
    f=(l_0^2 \zeta)^{1/3}\sinh^{2/3}\big(\pm \frac{3t}{2l_0}+\frac{b(R)}{l_0}\big).
\end{equation}
Another important limit is where $f\approx \frac{\zeta}{\omega}$. This is close to the region where the sign of $h=\dot{f}$ changes. To understand the behavior of $f$ in this region, we find it suitable to define $f=\big(\frac{\zeta}{d}\big)^{1/3}\hat{f}$ and choose $\omega=\big(\frac{\zeta}{d}\big)^{2/3}$ to find
\begin{equation}
    \hat{f}+\hat{f}\dot{\hat{f}}^2-\frac{\hat{f}^3}{l_0^2}=d.
\end{equation}
To study the region close to $f=\frac{\zeta}{\omega}$ we write $\hat{f}=d+\phi$ where $\phi\ll d$ and find (we assume $d\ll l_0$)
\begin{equation}
    \phi+d\,\dot{\phi}^2\approx\frac{d^3}{l_0^2},
\end{equation}
with solution
\begin{equation}
    \phi=\frac{d^3}{l_0^2}-\frac{d}{4}\big(\frac{t}{d}-l_0 c(R)\big)^2.
\end{equation}
Here $c(R)$ is an arbitrary function of $R$. We see that this has the same behavior as in Figure \ref{fig:ba} close to the crossover point $t_c=l_0 \,d\, c(R)$.

Given this intuitive picture, we proceed to study this geometry and its precise features.
Here we have the equation
\begin{equation}
   f \dot{f}^2=\zeta(R)-\omega(R)f+\frac{f^3}{l_0^2}
\end{equation}
The critical value of $f$ for which $\dot{f}$ becomes zero can be found by setting the left-hand side of this equation to zero. On the other hand, we require the only real solution to this equation to be a double zero point such that the derivative of the right-hand side of this equation with respect to $f$ vanishes at $f_c$.

Therefore we have 
\begin{equation}
    \zeta(R)-\omega(R)f_c+\frac{f_c^3}{l_0^2}=0, \qquad -\omega(R)+3\frac{f_c^2}{l_0^2}=0.
\end{equation}
From the second equality, we find
\begin{equation}
    f_c=\frac{l_0 \sqrt{\omega(R)}}{\sqrt{3}}\,,
\end{equation}
and using this in the first equation, we find
\begin{equation}
    \zeta(R)=\frac{2l_0}{3\sqrt{3}}\omega^{3/2}.
\end{equation}
This fixes $\omega(R)$ in terms of $\zeta(R)$.

Here, in contrast to the matter-dominated case, we do not have an upper bound on $f$. For $f<f_c$, we can choose the minus (black hole) branch, and for $f>f_c$, we can choose the plus (cosmological) branch.

\subsection{Curvature invariants at the apparent and event horizons}
It is shown in \cite{Modesto} that the curvature invariants at the radius $r=2M$ for McVittie's solution diverge. Here we find that the curvature invariants at the apparent horizon and also at $r=f=l_s$, where $l_s$ is the asymptotic horizon radius, for the cases studied in this section, remain finite over time. 

\begin{figure}
    \centering
    \includegraphics[width=0.9\linewidth]{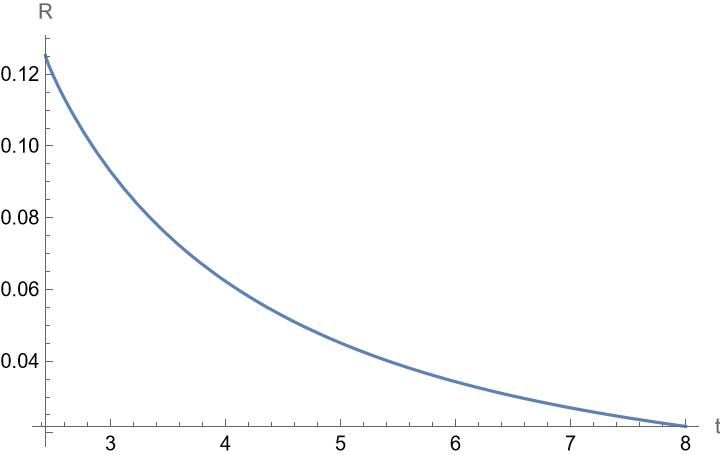}
    \includegraphics[width=0.9\linewidth]{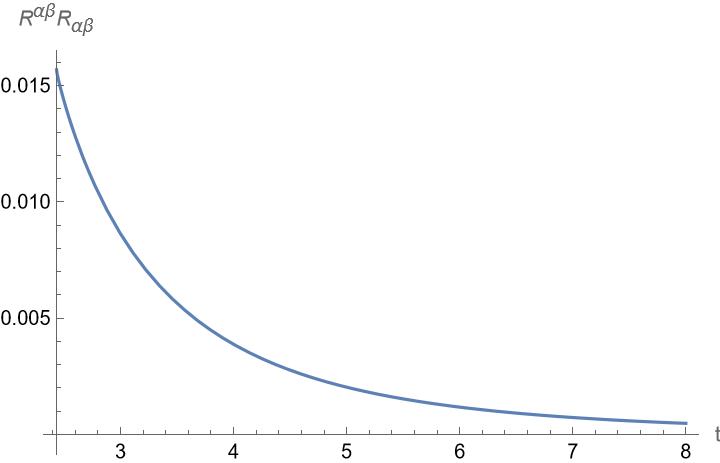}
    \includegraphics[width=0.9\linewidth]{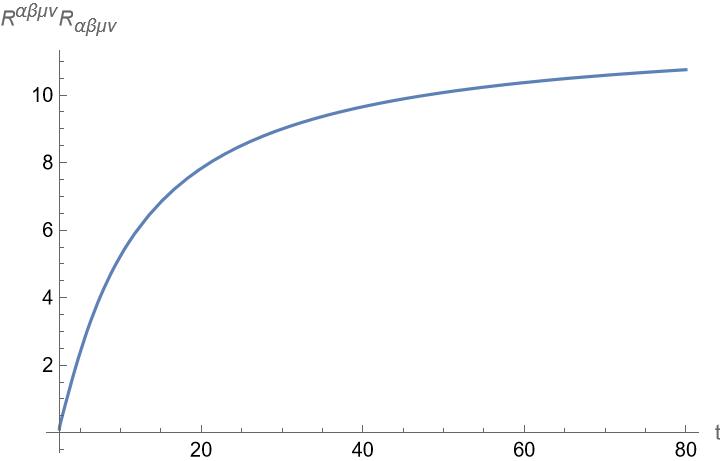}
    \caption{The evolution of curvature invariants at the apparent horizon over time}
    \label{fig:inv}
\end{figure}

At the location of the apparent horizon, we have
\begin{equation}
    h=\dot{f}=1.
\end{equation}
Considering the white hole case or, similarly, the near black hole region of the black hole case, we can write
\begin{equation}
    f=\zeta(R)^{\frac{1}{3}}\big(\frac{3t}{2}-b(R)\big)^{\frac{2}{3}},
\end{equation}
and at the horizon
\begin{equation}
    \dot{f}=\frac{\zeta^{\frac{1}{3}}}{\big(\frac{3t}{2}-b(R)\big)^{\frac{1}{3}}}=1.
\end{equation}
Therefore we have
\begin{equation}\label{constr}
    b(R_H)+\zeta(R_H)=\frac{3t}{2}.
\end{equation}
Replacing $t$ with $t_c=\frac{2}{3}(b(R_H)+\zeta(R_H))$ we can find
\begin{equation}
    \mathcal{R}=\frac{3\zeta'}{\zeta^2(-2b'+\zeta')},
\end{equation}
\begin{equation}
    \mathcal{R}^{\alpha\beta}\mathcal{R}_{\alpha\beta}=\frac{9\zeta'^2}{\zeta^4(-2b'+\zeta')^2},
\end{equation}
\begin{equation}
    \mathcal{R}^{\alpha\beta\mu\nu}\mathcal{R}_{\alpha\beta\mu\nu}=\frac{48\,b'^2+15\,\zeta'^2}{\zeta^4(-2b'+\zeta')^2}.
\end{equation}
We then find that with our choices $\zeta=l_s+\frac{\beta}{3}R^3$ and $b=\frac{\kappa l_s^4}{R^3}-d$ we have at $R\rightarrow 0$ \footnote{This follows from \eqref{rad2} and \eqref{rad3}.} for large $t$,
\begin{equation}
    \mathcal{R}=\mathcal{R}^{\alpha\beta}\mathcal{R}_{\alpha\beta}=0, \qquad \mathcal{R}^{\alpha\beta\mu\nu}\mathcal{R}_{\alpha\beta\mu\nu}=\frac{12}{l_s^4}.
\end{equation}
On the other hand, using \eqref{rad3} and \eqref{rht} we can sketch the evolution of these invariants over time,  see figure \ref{fig:inv}.

\begin{figure}
    \centering
    \includegraphics[width=0.9\linewidth]{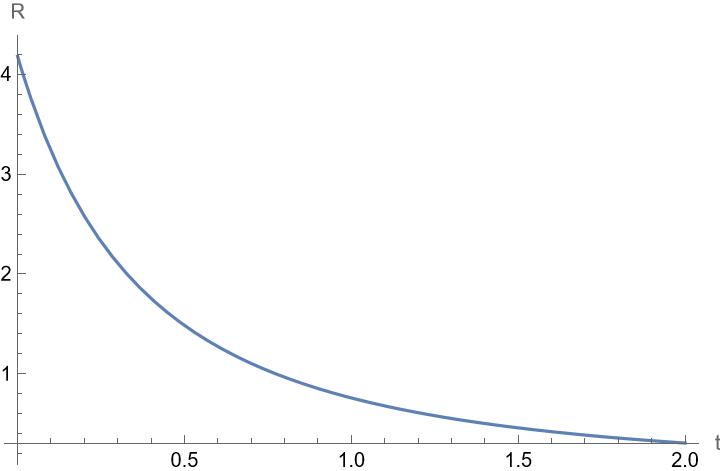}
    \includegraphics[width=0.9\linewidth]{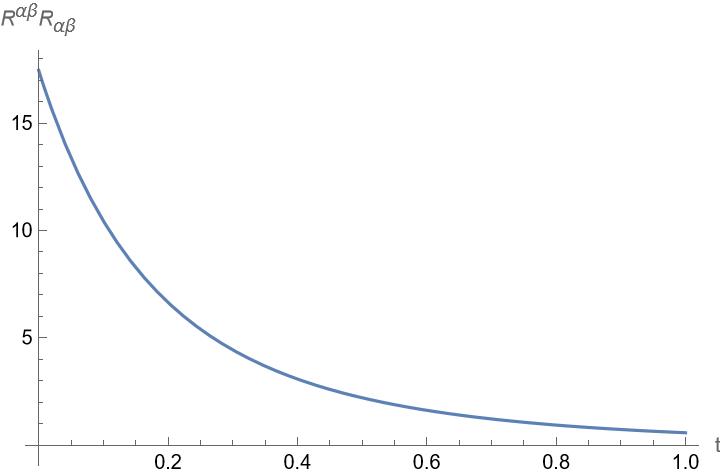}
     \includegraphics[width=0.9\linewidth]{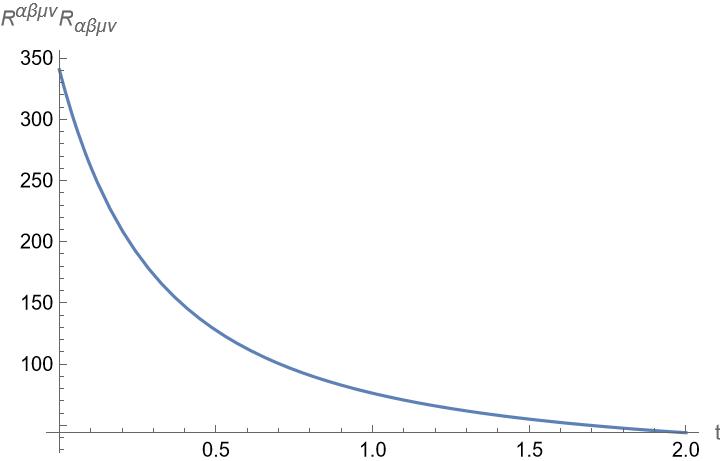}
    \caption{The evolution of curvature invariants at $f=l_s$ over time}
    \label{fig:invls}
\end{figure}

One can also consider the curvature invariants at the location of the asymptotic horizon $f=r=l_s$. We have sketched the results in Figure \ref{fig:invls}. We see that again the curvature invariants remain finite.

\section{Summary and Discussion}
Despite the recent claim that black hole horizons should be cosmologically coupled, in this work, we provide counterexamples for cosmological black hole metrics that have static or asymptotically static horizons. 

We find that it is technically much easier to find white hole metrics in an expanding cosmology or a black hole metric in a contracting cosmology. This is because the sign of $h$ in \eqref{ansatz} is negative for a black hole and positive for an expanding cosmology. Therefore, to find a black hole metric embedded in an expanding cosmology, we need to have a sign change in $h$ in the vicinity of the black hole relative to the asymptotic cosmology. 

Due to this complexity, we first studied the more trivial case of a white hole in expanding cosmology and then dealt with the issue of the sign change in $h$ by generalizing the form of the metric. The generalization is effectively a curvature-like behavior (relative to the flat FLRW metric).

As a summary, in the first part of this work, we proposed white and black hole metrics with static horizons embedded in FLRW cosmology. The metric has the following properties:

$a)$ It has a static horizon with no curvature singularity.

$b)$ It tends to the cosmological FLRW metric for large physical radius $r$.

$c)$ The energy density is the same as the cosmological case, but the pressure is different.

$d)$ The stress tensor also tends to the FLRW values for large $r$.

$e)$ It violates NEC. But we show that it is in principle possible to drop assumption c and find metrics with static horizons that satisfy NEC as well.

The corresponding stress tensor, assuming general relativity, violates the strong energy condition in the vicinity of the black hole.

$f)$ For the case of a constant Hubble parameter, it is equal to the Schwarzschild-de Sitter metric.

In \cite{Nojiri}, spherically symmetric solutions with possible static event horizons are proposed. We note that these constructions do not have asymptotic FLRW behavior. Using a redefinition of $t$, the general form of the metric, with a static event horizon in this case, is
\begin{equation}
    ds^2=-(1-\frac{r_0}{r})dt^2+\frac{b^2(t)}{1-\frac{r_0}{r}}dr^2+r^2 d\Omega_2^2,
\end{equation}
which does not have a cosmological FLRW behavior.

We also note that the Schwarzschild-de Sitter metric is not a counterexample to \cite{Faraoni}, since this metric can be written in a static patch.

In the second part of this work, we considered a more realistic framework with a constant equation of state (for the cases of matter and radiation) and first studied the white hole case due to its simplicity. We show that we have no non-trivial solutions for the case of $\omega\neq 0$. For the case of matter with $\omega=0$, we studied the white hole metric consistent with this assumption.  The crucial feature of this metric is that, despite having an asymptotically vanishing Hubble parameter, we have no singularities at the horizon.

We also studied the case of matter plus cosmological constant, where we assumed the pressure component remains equal to the constant cosmological value. In all of these cases, we show that the white hole horizon can asymptotically (for large $t$) tend to a static horizon.

Finally, we considered the case of a black hole immersed in cosmological matter (with and without cosmological constant) with a velocity function $h$ that changes sign in the vicinity of the black hole. We stress that in order to have a transition from an expanding cosmological phase to a black hole contracting phase, it is necessary to have a sign change in the velocity function. This sign change is only physically possible if we have a local positive curvature term $\omega>0$. Despite the common lore, this is not possible for other LTB-type metrics with vanishing cosmological constant, which have vanishing or negative $\omega$. In particular, for $\omega<0$ with $\zeta>0$ there is no critical value $f_c$ that allows such a transition. \\

\section*{Acknowledgment}
We thank Mohammad Mehdi Sheikh-Jabbari for comments and discussions. We specially thank Grigorii E. Volovik for mentioning his important work that helped us improve this manuscript.

\end{document}